\documentstyle [12pt] {article}

\newcommand {\be} {\begin{equation}}
\newcommand {\bea} {\begin{eqnarray}}
\newcommand {\ee} {\end{equation}}
\newcommand {\eea} {\end{eqnarray}}
\newcommand {\bi} {\bibitem}
\newcommand {\r} {\vec{r}}
\newcommand {\x} {\vec{x}}
\newcommand {\y} {\vec{y}}
\newcommand {\G} {\Gamma}
\newcommand {\p} {\psi}
\newcommand {\dd} {\tilde{\delta}}
\newcommand {\la} {\lambda}
\newcommand {\gd} {g^{\dagger}}
\newcommand {\ve} {V_e^{\dagger}}
\newcommand {\dve} {V_1^{\dagger}}

\newcommand {\iex} { e^{ \beta B(\x) \p (\la)}}

\newcommand {\es} { e^{ - \beta B s}}

\newcommand {\nor} {e^{\beta B (s-1)} + (s-1) e^{- \beta B}}
\newcommand {\norm} {1 + (s-1) \es}
\newcommand {\ie} {{\it i.~e.}}

\renewcommand{\theequation}{\thesection.\arabic{equation}}

\begin{document}
\title {Theory of continuum percolation II. Mean field theory}
\author{ Alon Drory}
\date{ \it Dipartimento di Fisica, Universit\'a {\sl La Sapienza}\\
        Piazzale Aldo Moro 2, Roma 00187, Italy}
\maketitle

\begin{abstract}
I use a previously introduced mapping between the continuum percolation model
and the Potts fluid to derive a mean field theory of continuum percolation
systems. This is done by introducing a new variational principle, the basis of
which has to be taken, for now, as heuristic. The critical exponents obtained 
are $\beta= 1$, $\gamma= 1$ and $\nu = 0.5$, which are identical with the mean
field exponents of lattice percolation. The critical density in this 
approximation is $\rho_c = 1/\ve$ where $\ve = \int d \x \, p(\x) \{ 
\exp [- v(\x)/kT] - 1 \}$. $p(\x)$ is the binding probability of two particles 
separated by $\x$ and $v(\x)$ is their interaction potential.
\end{abstract}

\newpage

\section{Introduction}
\setcounter{equation}{0}
 
The first paper in this series \cite{drory}, hereafter referred to as I, 
contained a general
formalism of continuum percolation. Such a system consists of classical
particles interacting through a pair potential $ v(\r_i, \r_j)$, such that they
can also bind (or connect) to each other with a probability $ p(\r_i,\r_j)$.
The function $ q(\r_i,\r_j) = 1 - p(\r_i,\r_j)$ is the probability of 
disconnection .

The formalism is based on a quantitative mapping between the percolation model
and an extension of the Potts model I have named the Potts fluid. For easy
reference, I recall here the essential definitions and results.

The $s$-state Potts fluid is a system of N classical spins $\{\la_i\}_{i=1}^N$
interacting with each other through a spin-dependent pair potential 
$V(\r_i,\la_i;\r_j,\la_j)$, such that
\be
V(\r_i,\la_i;\r_j,\la_j) \equiv V(i,j) = \left\{ \begin{array}{r @{\quad 
     \mbox{if} \quad}l}
   U(\r_i,\r_j) & \la_i =\la_j \\ W(\r_i,\r_j) & \la_i \neq\la_j 
   \end{array} \right.    \label {poten}
\ee
The spins are coupled to an external field $h(\r)$ through an interaction 
Hamiltonian 
\be
H_{int} = - \sum_{i=1}^N \,\p (\la_i) h(\r_i)
\ee
where
\be
\p (\la)  =  \left\{ \begin{array}{r @{\quad \mbox{if} \quad} l}
            s - 1 & \la = 1 \\ -1 & \la \ne 1 \end{array} \right. \label{psi}
\ee

Up to some unimportant constants, the Potts fluid partition function (more
precisely, the configuration integral) is
\be
Z = \frac{1}{N!} \sum_{\{\la_m\}} \int \, d\r_1 \cdots d\r_N \,
\exp \left[ -\beta \sum_{i>j} V(i,j) + \beta \sum_{i=1}^N h(i) \p (\la_i) 
\right] \label{conf}
\ee
The magnetization of the Potts fluid is defined as 
\be
M = \frac{1}{\beta N (s-1)}\, \frac{\partial \ln Z}{\partial h}
\label{defm}
\ee
where $h$ is the now constant external field. The susceptibility is
\be
\chi = \frac{\partial M}{\partial h}
\label{defchi}
\ee

The $n$-density functions of the Potts fluid, are defined as
\bea
\lefteqn{\rho^{(n)} (\r_1, \la_1; \r_2,\la_2; \ldots, \r_n, \la_n) =}
\nonumber \\
& &  \frac{1}{Z(N-n)!} \int d\r_{n+1} \cdots d\r_N \exp\left[ - \beta 
\sum_{i>j} V(i,j) - \beta \sum_{i=1}^N h(i) \p (\la_i) \right]
\nonumber \\
\label{defr}
\eea
Of particular interest is the spin pair-correlation function, defined as
\be
g_s^{(2)} (\x,\alpha; \y, \gamma) \equiv 
\frac{1}{\rho(\x) \rho(\y)}\, \rho^{(2)}(\x,\alpha;\y,\gamma) 
\label{pcorr}
\ee
which tends to $1$ when $| \x - \y| \to \infty $. Here $\rho(\x)$ is the 
numerical local density at the point $\x$. It is often useful to define 
a connected spin pair-correlation as
\be
h_s^{(2)}(\x,\alpha; \y, \gamma) \equiv g_s^{(2)} (\x,\alpha; 
\y, \gamma) - 1
\label{pccor}
\ee
This function tends to zero when $| \x - \y| \to \infty $.

Any continuum percolation model defined by $v(i,j)$ and $p(i,j)$ can be 
mapped onto an appropriate Potts fluid model with a pair-spin interaction 
defined by
\bea
U(i,j) & = & v(i,j) \nonumber\\
\exp \left[- \beta W(i,j) \right] & = & q(i,j) \exp \left[ - \beta v(i,j) 
\right]
\label{map}
\eea

The relation between the Potts magnetization and the percolation probability,
$P(\rho)$, is 
\be
\lim_{h \to 0} \, \lim_{N \to \infty} \, \lim_{s \to 1} \, M = P(\rho)
\label{prho}
\ee

For densities lower than the critical density, the susceptibility is 
directly related to the mean cluster size, $S$.
\be
\lim_{h \to 0} \,\lim_{N \to \infty} \, \lim_{s \to 1} \, \chi = \beta S
\qquad\qquad (\rho < \rho_c)
\label{chis}
\ee
 
An important quantity in the percolation model is the pair-connectedness 
function $\gd (\x, \y)$, the meaning of which is
\bea
\rho(\x)\rho(\y) \, \gd (\x, \y)\,\, d \x \, d \y &=& 
\mbox{Probability of finding two particles in regions } \nonumber \\
& & d \x \mbox{ and } d \y \mbox{ around the positions }
\x \mbox{ and } \y \mbox{, such} \nonumber \\
& & \mbox{that they both belong to the same cluster.}  \nonumber \\
\eea
This function is related to the mean cluster size by
\be
S = 1 + \rho\int d \r \, \gd (\r)
\label{Stwo}
\ee
where we assume, as we usually shall, that the system is translationally 
invariant, so that $\gd (\x, \y) = \gd (\x- \y)$ and 
$\rho(\x)= \rho(\y) = \rho$. 

The pair-connectedness is related to the Potts pair-correlation functions by
\bea
\gd (\x, \y) &=& \lim_{s \to 1} \left[ g_s^{(2)}( \x,\sigma ;
\y, \sigma) - g_s^{(2)}(\x,\sigma;\y, \eta) \right] \nonumber
\\
&=& \lim_{s \to 1} \left[ h_s^{(2)}( \x,\sigma ;\y, \sigma) - 
h_s^{(2)}(\x,\sigma;\y, \eta) \right]
\label{gdagf}
\eea
where $\sigma, \eta \ne 1$ and $\sigma \ne \eta$.

As usual, the first approach to any phase transition is to find the mean field
theory. This is the aim of the present paper. The mean field theory for the
magnetization (\ie, the percolation probability) and for the susceptibility
(\ie, the mean cluster size) is developed in section 2 from a physical point of
view. Then, in section 3, I show that the same results can be obtained from a
variational principle, which however, remains at present a heuristic device.
Section 4 uses the variational principle to obtain the pair-connectedness 
within the mean field approximation. Section 5 sums up the results.

\section{Mean field theory for $M$ and $\chi$}
\setcounter{equation}{0}

As usual, the basic assumption of mean field theory is that every spin $\la$ 
feels an average, homogeneous interaction $H(\la)/\beta$ due to all the other 
spins. Since the field $h$ distinguishes spins in state 1 from all the others,
we assume that $H(\la)$ takes two values, $H(\la=1)$, and $H(\la \ne 1)$, 
this last being identical for all spin states other than 1. This is because the
field preserves the symmetry between the non-1 spin states.

From Eq.~(\ref{conf}), the configuration integral (or the partition function, 
up to nonimportant constants) is, in this approximation,
\be
Z= \frac{1}{N !} \left\{ \int d\r \, \sum_{\la} \exp \left[ - H(\la) -\beta
h(\r) \p (\la) \right] \right\}^N \label{mfz}
\ee
Separating now the case $\la=1$ from the $s-1$ cases $\la \ne 1$, and recalling
the definition of $\p(\la)$, Eq.~(\ref{psi}), we have
\be
Z = \frac{1}{N !} \left[ \int d \r \,\, \Omega \right]^N 
\ee
with
\bea
\Omega &\equiv& \exp \left[ - H(\la=1) + \beta (s-1)h(\r)\right] \nonumber \\
&+&(s-1) \exp\left[ - H(\la \ne 1) - \beta h(\r) \right] \label{mfz2}
\eea

Let us denote
\bea
B_1 &\equiv& \exp\left[ - H(\la = 1 ) \right] \nonumber \\
B_r &\equiv& \exp\left[ - H(\la \ne 1 ) \right]
\eea
For a constant field, Eq.~(\ref{mfz2}) becomes
\be
Z = \frac{V^N}{N !} \left[ B_1 e^{\beta (s-1)h} + (s-1) B_r e^{-\beta h} 
\right]^N
\label{mfzf}
\ee
The magnetization is obtained from Eq.~(\ref{defm}),
\be
 M = \frac{1}{N(s-1) \beta} \, \frac{ \partial \ln Z}{\partial h} = 
\frac{B_1 e^{\beta (s-1)h} - B_r e^{-\beta h}}{B_1 e^{\beta (s-1)h} + 
(s-1) B_r e^{-\beta h}}
\label{mhere}
\ee
In the percolation limit, $ s \to 1$, we have
\be
M = 1 - \lim_{s \to 1}\left(\frac{B_r}{B_1}\right) e^{-\beta s h}   
\label{mfeq}
\ee
or, for $h=0$, when $M$ becomes the percolation probability $P$,
\be
P = 1 - \lim_{s \to 1}\left(\frac{B_r}{B_1}\right)  \label{pfeq}
\ee

This equation is a self-consistency condition because $(B_r/B_1)$ depends on
$M$. The reason is as follows. A spin $1$ interacts differently with other 
spins $1$ than it does with spins in other states. As a result, $B_1$ must 
depend on the average number of spins in state $1$ and of spins in all other 
states. The same reasoning holds for $B_r$. Now, let us denote
\bea
n_1 &\equiv& \mbox{ number of spins in state } 1 \nonumber \\
n &\equiv& \mbox{ number of spins in any state } \alpha \ne 1
\eea
I have shown in I [Eq.~(4.9)] that
\be
M = \frac{1}{N} \left( n_1 - n \right)
\ee 
Therefore $B_1$ and $B_r$ depend on $M$, (or $P$) through $n_1$ and $n$.

Finding the form of $(B_r/B_1)$ requires several approximate arguments, which 
will be clearer if we start with a simple case, \ie, $v(i,j) = 0$. In the 
percolation picture, this means no interactions . In the Potts fluid picture, 
this means that the only interactions take place between non-parallel spins, 
and these interactions are given by $\exp[-\beta W(i,j)] = q(i,j)$ [compare 
Eq.~(\ref{map})]. In this case, a spin $1$ interacts only with the $(s-1)n$ 
spins which are in non-1 states. Similarly, a non-1 spin interacts only with 
the $n_1 + (s-2)n$ spins which are non-parallel to it.

The first step is to notice that here we can take already the limit $s \to 1$,
since the term $(s-1)$ in the denominator in the definition of $M$, 
Eq.~(\ref{defm}), has vanished in Eq.~(\ref{mfeq}). In this limit, a spin $1$ 
interacts with zero [the limit of $(s-1)n$] other spins, while any non-1 spin 
interacts with $n_1 - n = N M$ other spins.

The main argument is now that it is easier to find $B_r$ and $B_1$ in the limit
$M \to 1$, \ie, when practically all the spins are in the state $1$.

Consider first a state where all the spins are in the state $1$. Such a state 
can be thought of, for example, as the limit of an infinitely strong field. 
Then there are no inter-spin interaction, and the partition function is
\be
Z = \frac{1}{N!} \int d1 \cdots dN \exp[\beta N h (s-1)] = \frac{V^N}{N!}
\exp[\beta N h (s-1)] \label{z1}
\ee
However, by definition, we also have
\be
Z = \frac{V^N}{N!}B_1^N \exp[\beta N h (s-1)] \label{z2}
\ee
Comparing the equations, we have that $B_1 = 1$. In principle, this holds in 
the limit $M \to 1$ only; however, as mentioned above, in the $s\to 1$ limit, 
a spin $1$ interacts with no other non-1 spins. On the other hand, in our case,
the interaction with other spins $1$ is $v(i,j)=0$. Therefore, {\it all} the 
interactions vanish and the above result for $B_1$ can be extended to all 
values of $M$. Hence, $B_1 = 1$ in general.

Next consider a state where all spins but one are in the state $1$ (in the 
thermodynamic limit, this is still $M=1$ ). Again, the partition function can 
be calculated easily. For definiteness, assume that the non-1 spin is always 
numbered $N$. Then, from Eq.~(\ref{conf}), we have
\bea
Z &=& \frac{1}{(N-1)!}\int \, d1 \cdots dN \,
\exp \left\{ -\beta \sum_{i=1}^{N-1} W(i,N) + \beta h [(N-1)s -N] \right\}
\nonumber \\
&=&\frac{1}{(N-1)!}\left[\int d\r_i \, q(\r_N - \r_i)\right]^{N-1} \exp\left\{
 \beta h [(N-1)s -N]\right\}
\eea
And we must also have
\be
Z = \frac{V^N}{(N-1)!}B_r B_1^{N-1} \exp\left\{\beta h [(N-1)(s-1) -1]\right\}
\ee
Hence,
\be
B_r = \left[ \frac{1}{V} \int d\r \, q(\r) \right]^{N-1}
\ee
Using $p(\r) \equiv 1 - q(\r)$ and denoting
\be
V_e \equiv \int d\r \, p(\r) \>, \label{ve}
\ee 
we have
\be
B_r = \left[ 1 - \frac{V_e}{V}\right]^{N-1}
\ee
Now the power $N-1$ represents the number of spins which interact with the 
non-1 spin. In other words, it is $n_1 - n$ for this particular configuration.
In the thermodynamic limit it can therefore be replaced by $M N$. In terms of
the density $\rho = N/V$ we can write
\be
B_r = \left[ 1 - \frac{\rho V_e}{N}\right]^{M N} \label{br}
\ee
Now although this equation was derived for the case $M=1$, it is actually 
written for an arbitrary $M$. Consequently, we'll {\it assume} now that we can
use it for all values of $M$. This is part of the mean field approximation 
itself, \ie, that for the case of $v(i,j)=0$, we suppose that in the 
thermodynamic limit,
\be
\frac{B_r}{B_1} = \lim_{N\to \infty} \left[ 1 - \frac{\rho V_e}{N}\right]^{M N}
= \exp\left(-M \rho V_e \right) \label{br1}
\ee
Substituting this into Eq.~(\ref{mfeq}) yields
\be
M = 1 - \exp\left(-M \rho V_e  - \beta s h\right) \label{mfm}
\ee
or, for $h = 0$, 
\be
M = 1 - \exp\left(-M \rho V_e \right) \label{mfm2}
\ee
This equation is very similar to the one obtained in the mean field theory of 
the Ising model. For low densities, its only solution is $M=0$. A non-zero
solution appears first when the slopes of the two sides of the equation are 
equal at $M=0$. This condition yields for the critical density, $\rho_c$, the 
result
\be
\rho_c = \frac{1}{V_e} \label{rhoc}
\ee
Typically \cite{pike}, the function $p(\r)$ is chosen to be
\be
p(\r) = \left\{ \begin{array}{r @{\quad \quad}l}
   1 & \r \quad \mbox{  belongs to some excluded volume,}\, V_{exc},\\
 
     &    \mbox{\qquad around the center of the particle} \\ 
   0 & \mbox{\qquad else} 
\end{array}\right.
\ee
Then, $V_e = V_{exc}$, and Eq.~(\ref{rhoc}) becomes
\be
\rho_c = \frac{1}{V_{exc}} 
\ee
Computer simulations \cite{alon} show this result to be correct in the limit of
infinite spatial dimension. This is precisely where mean field theory is 
expected to be correct.

Expanding now Eq.~(\ref{mfm}) in the vicinity of the critical point, we have
\be
M = M\frac{\rho}{\rho_c} - \frac{1}{2}M^2\left(\frac{\rho}{\rho_c}\right)^2 + 
\ldots
\ee
Hence, the percolation probability, which is equal to $M$ in this limit, is
\be
P(\rho) = M \approx \frac{2\rho_c}{\rho^2} \left(\rho_c - \rho \right) 
\sim  \left(\rho_c - \rho \right)^{\beta} \qquad (\rho \to \rho_c)
\ee
where in this equation, $\beta$ is the critical exponent of $P(\rho)$.
Therefore, we find
\be
\beta = 1 \qquad \mbox{(Mean Field)} \label{beta}
\ee

We can now calculate the susceptibility. From Eq.~(\ref{mfm}),
\be
\chi = \frac{\partial M}{\partial h} \bigg\vert_{h = 0} = \left[ \beta + 
\rho V_e
\frac{\partial M}{\partial h} \bigg\vert_{h = 0} \right]\exp\left(-M \rho V_e
\right)
\ee
or
\be
\chi = \frac{\beta e^{- M \rho V_e}}{1 - \rho V_e}
\ee
For $\rho < \rho_c$, we have that $\chi \to \beta S$. In this case, $M=0$, so 
that the mean cluster size is
\be
\frac{\chi}{\beta} \to S = \frac{1}{1 - \rho V_e} \sim (\rho_c -\rho)^{-\gamma}
\ee
Hence, the critical exponent $\gamma$ is
\be
\gamma = 1 \qquad \mbox{(Mean Field)} \label{gamma}
\ee

We now wish to extend this argument to cases where  $v(i,j)\ne 0$. The 
essentials remain unchanged. Again we look at the limit $M \to 1$. This time, 
when all the spins are in the  state $1$, the equivalents of Eqs.~(\ref{z1})and
(\ref{z2}) yield
\be
B_1^N = \frac{1}{V^N} \int d1 \cdots dN \exp\left[ - \beta \sum_{i=1}^N
\sum_{i > j} v(i,j) \right]  \label{bone}
\ee
Similarly, if only one spin is in a state other than $1$, we have
\bea
B_1^{N-1}B_r = \frac{1}{V^N} \int d1 \cdots dN & &\exp\left[ - \beta 
\sum_{i=1}^{N-1}\sum_{i > j} v(i,j) \right] \nonumber \\
& &\times \prod_{i=1}^{N-1} q(i,N)\exp[-\beta v(i,N)]
 \label{btwo}
\eea
Dividing the two equations, we find that in this limit (\ie, $M \to 1$)
\be
\frac{B_r}{B_1} = \frac{\int d1 \cdots dN \, \prod\limits_{i=1}^{N-1} q(i,N)
\exp \left[- \beta \sum\limits_{i>j=1}^{N-1}v(i,j)-\beta v(i,N)\right]}
{\int d1 \cdots dN \, \exp \left[- \beta \sum\limits_{i>j=1}^N v(i,j)\right]}
\label{brb1}
\ee
In general, one cannot compute exactly this ratio. However, we don't need 
$B_r / B_1$ exactly, but only in the region $M \approx 0$, where
the critical behavior occurs. Now, from the previous arguments, $B_r / B_1$ 
can depend on $M$ only through the excess spin-1 density $M\rho = \rho_1 - 
\rho_{\alpha}$ (where $\rho_1$ is the density of spins $1$ and $\rho_{\alpha}$
is the density of spins in any state $\alpha \ne 1$). Hence, we
are interested in the region of small $M\rho$. At first sight, this seems 
unhelpful, because Eq.~(\ref{brb1}) corresponds to the limit $M \to 1$. 
However, $M\rho$ may be made small by decreasing the density rather than $M$.
In other words, the region of interest $M\rho \to 0$ may be obtained by looking
at the limit $\rho \to 0$ even if $M \to 1$.

We therefore take the small density limit of Eq.~(\ref{brb1}) and extract the 
leading behavior. Let us introduce the functions
\bea
f(\r) &=& \exp[-\beta v(\r)] - 1   \nonumber \\
f^{*}(\r) &=& q(\r)\exp[-\beta v(\r)] - 1  \nonumber \\
f^{\dagger}(\r) &=& f(\r) - f^{*}(\r) = p(\r)\exp[-\beta v(\r)]
\eea
The function $f(\r)$ is just the Mayer f-function \cite{hansen}. Its usefulness
stems from its being typically short ranged (\ie, $f(\r) \to 0$ at $\vert \r 
\vert \to \infty$). Therefore, integrals over $f$ (and over $f^{*}$ and 
$f^{\dagger}$ as well) remain finite in the thermodynamic limit.

Eq.~(\ref{brb1}) now becomes
\be
\frac{B_r}{B_1} = \frac{\int d1 \cdots dN \prod\limits_{i=1}^{N-1}\left\{ 
\prod\limits_{i>j} \left[1 + f(i,j)\right]\left[1+ f^*(i,N)\right]\right\}}
{\int d1 \cdots dN \prod\limits_{i=1}^{N-1} \prod\limits_{i>j} 
\left[1 + f(i,j)\right]}
\ee
Expanding this expression, we find
\bea
\frac{B_r}{B_1} &=& \frac{\int d1 \cdots dN \left\{ 1 + \sum\limits_{i=1}^
{N-1}\left[\sum\limits_{i>j}f(i,j) + f^*(i,N)\right] + \cdots \right\}}
{\int d1 \cdots dN \left\{ 1 + \sum\limits_{i=1}^{N-1}\sum\limits_{i>j}
f(i,j)+\cdots \right\}} \nonumber \\
&=&\frac{1 + [(N-1)(N-2)/2V]\int d\r \, f(\r) + (N-1/V)\int d\r \,
f^*(\r) + \cdots} {1 + [(N-1)(N-2)/2V]\int d\r \, f(\r) + (N-1/V)
\int d\r \, f(\r) + \cdots} \nonumber \\
\eea
Keeping now only the leading terms in the density, we obtain
\be
\frac{B_r}{B_1}\approx \frac{1 + (N-1/V)\int d\r \,f^*(\r) + \cdots} {1 + 
(N-1/V) \int d\r \, f(\r) + \cdots} = 1 - \frac{N-1}{V} \int d\r \,
f^{\dagger} (\r) + \cdots
\ee
where we have used the definition $f^{\dagger}(\r)=f(\r) - f^{*}(\r)$. Again, 
as in Eq.~(\ref{br}), we recognize that the term $N-1$ represents just $n_1 -n$
in this configuration. Hence once again, we can replace it with $M N$ and write
\be
\frac{B_r}{B_1}\approx 1 - \frac{M N}{V} \int d\r \, f^{\dagger} (\r) + \cdots
= 1 - M \rho V_e^{\dagger} \label{bfinal}
\ee
where 
\be
V_e^{\dagger} \equiv \int d\r  \, f^{\dagger} (\r)
\label{vedef}
\ee
generalizes Eq.~(\ref{vedef}), to which it properly reduces if $v(\r) = 0$.

To the same order in $\rho$, we may also express Eq.~(\ref{bfinal}) as
\be
\frac{B_r}{B_1}\approx \exp\left[- M \rho V_e^{\dagger}\right] \label{bfin}
\ee
which completes the analogy with Eq.~(\ref{br1}) for the case $v=0$. Because of
this, all the results for the case $v=0$ apply to the more general case as 
well. Hence, for arbitrary potentials and binding criteria, we now find
\bea
P &=& 1 - \exp\left(- \rho P \ve \right) \nonumber \\
S &=& \frac{1}{1 - \rho \ve} \qquad\qquad (\rho < \rho_c) \nonumber \\
\rho_c &=& \frac{1}{\ve} = \frac{1}{\int d\r \, f^{\dagger}(\r)} 
\nonumber \\
\beta_{MF} &=& 1 \nonumber \\
\gamma_{MF}&=&1 \label{mfres}
\eea

\section{ A heuristic variational formulation}
\setcounter{equation}{0}

We would like now to calculate the connectedness function in the mean field 
approximation. The previous arguments are inadequate for this task.
One knows, however, that in the Ising model, the correlation function may be 
calculated within mean field theory through a variational formulation based on
the Bogolyubov-Feynman inequality \cite{binney}, \ie,
\be
\frac{Z}{Z_0} \ge \exp\left(-\beta \langle {\cal{H}}_1 \rangle_0 \right)
\ee
where the subscript $0$ refers to some reference system and ${\cal{H}}_1 = 
{\cal{H}} - {\cal{H}}_0$, where ${\cal{H}}$ is the true Hamiltonian of 
the system and ${\cal{H}}_0$ is the Hamiltonian of the reference system.

Unfortunately, this inequality is inadequate to deal with our system, because
the potential $v(i,j)$ will typically have a strongly repulsive part (``hard 
core'') at short distances, or else the function $q(i,j)$ will typically vanish
in some range (where the binding is certain), both of which make the r.h.s of 
the inequality undetermined. One would therefore like a variational formulation
of the mean field which allows for strong effective interactions. I will show
that the mean field developed in the previous section can be derived from such 
a principle, though I have been unable to formally prove the principle itself. 
Its status must therefore be taken at present to be essentially heuristic.

The motivation is as follows. Let us assume, as usual, some reference system in
which the spins interact only with an external field $B$. The reference 
Hamiltonian, ${\cal{H}}_0$, is 
\be
{\cal{H}}_0 = - B \sum_{i=1}^N \p (\la_i)
\ee
Then the partition function can be written as
\bea
Z &=& \frac{1}{N!} \sum_{\{\la_m\}} \int \, d\r_1 \cdots d\r_N \,
\left\{\exp \left[ -\beta \sum_{i>j} V(i,j) + \beta (h -B)\sum_{i=1}^N 
\p (\la_i) \right] \right\} \nonumber \\
& & \qquad\qquad \qquad\qquad\qquad \times \exp\left[\beta B\sum_{i=1}^N 
\p (\la_i) \right]   \nonumber \\
&=&Z_0 \left\langle \exp \left[ -\beta \sum_{i>j} V(i,j) + \beta (h -B)
\sum_{i=1}^N \p (\la_i) \right] \right\rangle_0
\eea
where
\bea
Z_0 &=& \frac{1}{N!} \sum_{\{\la_m\}} \int \, d\r_1 \cdots d\r_N \,
\exp\left[\beta B\sum_{i=1}^N \p (\la_i)\right] \nonumber \\
&=& \frac{V^N}{N!}\left[e^{\beta B (s-1)} + (s-1) e ^{- \beta B} \right]^N 
\label{zo}
\eea
and $\langle \quad \rangle_0$ means an average performed in the reference 
system with the Hamiltonian ${\cal{H}}_0$. The convexity of the exponential 
function then gives us the Bogolyubov-Feynman inequality,
\be
\frac{Z}{Z_0} \ge \exp\left[ \left\langle -\beta \sum_{i>j} V(i,j) + \beta 
(h -B)\sum_{i=1}^N \p (\la_i) \right\rangle_0  \right] \label{bog}
\ee
The usual mean field theory follows from extremization with respect to the
parameter $B$. The resulting approximation to $Z$ is expected to be good as 
long as $\beta V$ is small. Now, if this is the case, we could as well write
\be
-\beta V(i,j) \approx \exp\left[-\beta V(i,j)\right] -1 \label{rep}
\ee
so that to this order
\be
\frac{Z}{Z_0} \sim \exp\left[ \sum_{i>j} \left\langle  e^{-\beta V(i,j)} - 1
\right\rangle_0 + \beta (h -B)\sum_{i=1}^N \left\langle \p (\la_i) \right
\rangle_0  \right] \label{approx}
\ee
However, at this stage it is not clear what has become of the inequality in 
Eq.~(\ref{bog}), because the replacement Eq.~(\ref{rep}) is increasing the
exponent, thereby countering the Bogolyubov-Feynman inequality.

Nevertheless, for small $\beta V$, we expect that $(1/N)\ln Z $ can be 
approximated  by the following function,
\be
F_t(B) \equiv \frac{1}{N}\left\{ \ln Z_0 + \sum_{i>j} \left\langle  \exp\left[
-\beta V(i,j)\right] - 1\right\rangle_0 + \sum_{i=1}^N \beta (h -B) \left
\langle \p (\la_i) \right\rangle_0 \right\}  \label{ft}
\ee
Typically, of course, $V(i,j)$ is not uniformly small. The surprising result,
however, is that if we use for $(1/N)\ln Z$ the {\it extremum} value of 
$F_t(B)$ with respect to $B$, we shall recover the mean field theory of the 
previous section.

The proof is straightforward. First,
\be
\left\langle \p (\la_i) \right\rangle_0 = \frac{\sum_{\la_i} \p (\la_i)
\exp \left[\beta B \p (\la_i)\right]}{\sum_{\la_i} \exp \left[\beta 
B \p (\la_i)\right]} = \frac{(s-1)\left[e^{\beta B (s-1)} - e^{- \beta B}
\right]}{\nor}
\label{psimf}
\ee
Next, we rewrite
\bea
\exp\left[- \beta V(i,\la_i;j,\la_j)\right] - 1 &=& q(i,j)\exp[-\beta v(i,j)] 
- 1 \nonumber \\
&+& p(i,j)\exp[-\beta v(i,j)]\delta_{\la_i,\la_j}
\eea
Taking the average with respect to the reference system of the two terms on the
r.h.s, we find first
\be
\left\langle q(i,j)\exp[-\beta v(i,j)] - 1 \right\rangle_0 = \frac{1}{V^2}
\int d i \, d j \,\left\{ q(i,j)\exp[-\beta v(i,j)] - 1 \right\}
\ee
which is independent of B. 

The second term is
\bea
\lefteqn{\left\langle p(i,j)\exp[-\beta v(i,j)]\delta_{\la_i,\la_j} \right
\rangle_0} \nonumber \\
 &=&\frac{1}{Z_0 N!} \int d1 \cdots dN \sum_{\{\la_m\}} p(i,j)\exp\left[ -\beta
v(i,j) + \beta B \sum_{k=1}^N \p (\la_k) \right]\delta_{\la_i,\la_j} \nonumber
\\
&=& \left[\frac{1}{V^2}\int d i \, d j \, \, p(i,j)e^{- \beta v(i,j)} \right]
\frac{\sum\limits_{\la_i,\la_j} \, \delta_{\la_i,\la_j} \exp\left\{\beta B 
[\p (\la_i) + \p (\la_j)]\right\} }{\left[\nor \right]^2}\nonumber \\
&=&\left[\frac{1}{V^2}\int d i \, d j \, \, p(i,j)e^{- \beta v(i,j)} \right] 
\frac{e^{2 \beta B (s-1)} + (s-1)e^{- 2 \beta B}}{\left[\nor\right]^2}
\label{vimf}
\eea
Note that from the definition of $\ve$, Eq.~(\ref{ve}), we have
\be
\frac{1}{V^2} \int d i \,d j \,\,p(i,j)\exp[- \beta v(i,j)] = \frac{\ve}{V}
\ee

Substituting Eqs.~(\ref{zo}), (\ref{psimf}) and (\ref{vimf}) into the 
definition of $F_t(B)$, Eq.~(\ref{ft}), we have
\bea
F_t(B) &=& C + \ln\left[\nor\right] \nonumber \\
&+ & \frac{N-1}{V} \left(\frac{\ve}{V}\right)\frac{e^{2 \beta B (s-1)} 
+ (s-1)e^{- 2 \beta B}}{\left[\nor\right]^2} \nonumber \\
&+ &  \beta (h - B)(s-1)\frac{e^{\beta B (s-1)} - e^{- \beta B}}{\nor}
\label{ftb}
\eea
where
\be
C = \left\langle q(i,j)\exp[-\beta v(i,j)] - 1 \right\rangle_0 + \ln V - \frac
{1}{N}\ln(N!)
\ee
is independent of $B$.

We now have immediately that
\be
\frac{\partial F_t}{\partial B} = \frac{\beta^2 s^2 (s-1)(h-B) \es}{\left[
\norm\right]^2} + \left[\frac{(N-1)\ve}{V}\right]\frac{\beta s (s-1) \es 
\left( 1 - \es \right)}{\left[\norm\right]^3}
\label{derf}
\ee

The extremum condition $\partial F_t / \partial B = 0$ yields the equation
\be
B - h = \left(\frac{\rho \ve}{\beta s}\right) \frac{1 - \es}{\norm}
\label{condmf}
\ee
In this approximation, the magnetization is
\be
M = \frac{1}{\beta (s-1)}\frac{\partial F_t}{\partial h}=\frac{1}{\beta (s-1)}
\left[\frac{\partial F_t}{\partial B}\frac{\partial B}{\partial h} +
\left(\frac{\partial F_t}{\partial h}\right)_B \right] = \frac{1}{\beta (s-1)}
\left(\frac{\partial F_t}{\partial h}\right)_B
\ee
where we have used the extremum condition $\partial F_t / \partial B = 0$.

From Eqs.~(\ref{ftb}) and (\ref{condmf}), we have
\be
M = \frac{e^{\beta B (s-1)} - e^{- \beta B}}{\nor} = \frac{\beta s}{\rho \ve}
\left( B - h \right)
\label{mb}
\ee
Hence we can rewrite Eq.~(\ref{condmf}) as
\be
\beta s B = \beta s h + M \rho \ve
\label{bcond}
\ee
Substituting this result back into Eq.~(\ref{mb}) gives us
\be
M = \frac{1 - e^{- M \rho \ve} e^{- \beta s h}}{1 + (s-1) e^{- M \rho \ve} 
e^{- \beta s h}}
\ee
which is exactly the mean field equation, Eq.~(\ref{mhere}) with $B_r / B_1$ 
given by Eq.~(\ref{bfin}). Hence, indeed, all the mean field results may be 
obtained from the variational principle of $F_t$.

Finally, note that from Eq.~(\ref{derf}), the extremum of $F_t$ is actually a 
maximum. This might suggest the existence of an inequality $\ln Z \ge N F_t$, 
but I have been unable to prove it one way or another.

\section{Correlation functions}
\setcounter{equation}{0}

The variational principle now allows us to find the correlation function in
mean field theory. To this end, we need to relate the Potts $n$-density
functions defined in Eq.~(\ref{defr}) to the partition function. By analogy 
with a similar formalism in the theory of liquids \cite{hansen}, let us define
a generalized functional differentiation operator, in the following way. Let 
${\cal{F}}[t(\r, \la)]$ be a functional of the function $t(\r, \la)$, which 
depends on a position variable as well as on an associated discrete spin 
variable. Then the generalized functional derivative 
$\dd {\cal{F}} / \delta t$ will be defined through the relation
\be
\delta {\cal{F}}=\int d \r \,\sum_{\la} \frac{\dd {\cal{F}}}{\delta t(\r,\la)}
\, \delta t(\r, \la)
\label{deffd}
\ee
where $\delta {\cal{F}}$ is the change in ${\cal{F}}$ associated with a 
variation $\delta t$ in $ t(\r, \la)$. The only difference with the usual 
functional derivative is in the added summation over the spin variable. This 
does not change any of the basic properties of the operator.

I shall show later that the Potts correlation functions are functional 
derivatives of the partition function, for fields $h$ which are inhomogeneous.
In order to extend the mean field theory to cover this case, we now need to 
adopt the following heuristic claim:

{\bf Heuristic Assumption}: The $F_t$ variational principle yields the proper 
mean field theory of continuum percolation even for the case of inhomogeneous 
fields $h(\r)$ and $B(\r)$. The extremum condition must then be generalized to
be
\be
\frac{\dd F_t}{\delta \left[\beta B(\r)\p(\la)\right]} = 0
\label{extr}
\ee
which determines the extremizing field $B(\r)$.

The application of this extremum condition yields a complicated integral
equation. However, for our purpose, we only need $F_t[B]$ in the region $h(\r)
\to 0$ and in the vicinity of the critical region $M(\r) \to 0$. Looking at the
results for the homogeneous case, Eq.~(\ref{bcond}), we can expect that the 
extremizing function $B(\r)$ will be also very small in this case (the solution
shows this assumption to be self-consistent). Therefore, one only needs to look
at the extremum condition for very small $B(\r)$. The details of this 
calculation are in the Appendix. The final result is the condition (to first 
order in $B$)
\be
B(\x) = h(\x) + \frac{\rho}{s}\int d \y \,\,p(\x- \y)
e^{-\beta v(\x-\y)} B(\y)
\label{exb}
\ee
where we have assumed that $p(\x,\y)= p(\x- \y)$. Note that if $B$ and $h$ are
uniform, we recover the first order limit of Eq.~(\ref{condmf}), as we should.
This equation is easily solved in Fourier space where
\be
\hat{B}(\vec{k}) = \frac{\hat{h}(\vec{k})}{1 - (\rho/s) \hat{f^{\dagger}}
(\vec{k})}
\label{bfs}
\ee
with
\be
f^{\dagger}(\r) = p(\r)e^{-\beta v(\r)} 
\label{exft}
\ee
and functions with hats are Fourier transforms of the corresponding functions 
without hats. 

Let us now calculate the Potts correlation functions. We take
\be
t(\r, \la) = \exp[\beta h(\r) \p (\la)] \quad .
\label{deft}
\ee
Then we have for the partition function from Eq.~(\ref{conf})
\be
Z = \frac{1}{N!} \sum_{\{\la_m\}} \int \, d\r_1 \cdots d\r_N \,
\prod_{i=1}^N t(\r_i, \la_i) e^{-\beta \sum_{i>j} V(i,j)}
\label{part}
\ee
It is now simple to see that
\be
\frac{\dd Z}{\delta t(\r, \la)} = \frac{N}{N!} \sum_{\{\la_2,\ldots, \la_N\}} 
\int \, d\r_2 \cdots d\r_N \,\prod_{i=2}^N t(\r_i, \la_i)\left[e^{-\beta 
\sum_{i>j} V(i,j)}\right]_{{\r_1 = \r} \atop 
{\la_1 = \la}}
\ee
Comparing this to Eq.~(\ref{defr}) we see that
\be
\rho^{(1)} (\r, \la) = \frac{t(\r, \la)}{Z} \frac{\dd Z}{\delta t(\r, \la)} =
\frac{\dd \ln Z}{\delta \ln t(\r, \la)} \label{ro1}
\ee
Similarly, we have that
\bea
\lefteqn{\frac{\dd^{(2)} Z}{\delta t(\x, \alpha)\delta t(\y, \gamma)}}
\nonumber \\
&=& \frac{1}{(N-2)!} \sum_{\{\la_3,\ldots, \la_N\}} 
\int \, d\r_3 \cdots d\r_N \,\prod_{i=3}^N t(\r_i, \la_i)\left[e^{-\beta 
\sum_{i>j} V(i,j)}\right]_{{\r_1 = \x; \,\la_1 = \alpha}
\atop {\r_2 = \y;\,\la_2 = \gamma}} \nonumber \\
\eea
And a comparison with Eq.~(\ref{defr}) shows that
\be
\rho^{(2)} (\x,\alpha;\y,\gamma) - \rho^{(1)} (\x,\alpha)\rho^{(1)}(\y,\gamma)
 = t(\x, \alpha)t(\y, \gamma)\frac{\dd^2 \ln Z}{\delta t(\x, \alpha)
\delta t(\y, \gamma)} 
\label{ro2}
\ee

Within the mean field theory we replace $\ln Z$ with $N F_t[B]$. Making use of
the extremum condition, we now have from Eq.~(\ref{ro1}) and the definition
of $t(\r, \la)$, Eq.~(\ref{deft}), that
\be
\rho^{(1)} (\x, \alpha) = \left[\frac{\dd N F_t}{\delta [\beta h(\x) \p 
(\alpha)]}\right]_B = \frac{N}{\G}  \exp\left[\beta B(\x) \p (\alpha)\right] 
\ee
where
\be
\G \equiv \sum_{\la} \int d \vec{z}\,\exp \left[\beta B(\vec{z}) \p(\la)\right]
\label{ro1h}
\ee
The easiest way to calculate the pair-correlation function is now to use the
identity
\bea
\frac{\dd \rho^{(1)}(\x, \alpha)}{\delta [\beta h(\y) \p (\gamma)]} &=& 
t(\y, \gamma)\frac{\dd}{\delta t(\y, \gamma)} \left[ t(\x, \alpha)
\frac{\dd N F_t}{\delta t(\x, \alpha)}\right] \nonumber \\
&=& t(\y,\gamma)\frac{\dd N F_t}{\delta t(\x, \alpha)} \delta (\x - \y) \,
\delta_{\alpha, \gamma} + t(\x, \alpha)t(\y, \gamma)\frac{\dd^2 N F_t}
{\delta t(\x, \alpha)\delta t(\y, \gamma)} \nonumber \\
&=& \rho^{(1)}(\x, \alpha) \delta (\x - \y) \,\delta_{\alpha, \gamma} + 
\rho^{(1)}(\x, \alpha)\rho^{(1)}(\y, \gamma)h^{(2)}(\x,\alpha;\y,\gamma)
\nonumber \\
\label{idrho}
\eea
On the other hand, we have from Eq.~(\ref{ro1h}) that
\be
\frac{\dd \rho^{(1)}(\x, \alpha)}{\delta [\beta h(\y) \p (\gamma)]} = 
\left[ 1 - \frac{1}{\G}e^{\beta B(\y) \p (\gamma)}\right] 
\frac{N}{\G} e^{\beta B(\x) \p (\alpha)} \frac{\dd [B(\x)\p(\alpha)]}
{\delta [h(\y) \p (\gamma)]}
\label{eigh}
\ee

We are interested in this function in the range $\rho < \rho_c$ and $h = 0$, 
which is the one relevant for percolation. In this case, both $h$ and $B$ 
vanish, so that we can evaluate all quantities at zero fields. Thus, using 
the fact that $\G = s V$ in this limit (see the Appendix), we have that
\be
\rho^{(1)} (\x, \alpha)\bigg\vert_{B=0} =\frac{N}{\G}\bigg\vert_{B=0} 
= \frac{\rho}
{s}  \label{r1fin}
\ee
where $\rho = N/V$ as usual. In the same limit, we also have from 
Eq.~(\ref{eigh})
\be
\frac{\dd \rho^{(1)}(\x, \alpha)}{\delta [\beta h(\y) \p (\gamma)]}
\bigg\vert_{h=0,B=0} =  \frac{\rho}{s} \left[ 1 - \frac{1}{s V}\right]
\frac{\dd [B(\x)\p(\alpha)]}{\delta [h(\y) \p (\gamma)]}\Bigg\vert_{h=0, B=0}
\label{rhofin}
\ee
In the thermodynamic limit, the second term inside the brackets on the r.h.s 
vanishes (its presence is typical of the use of the canonical rather than the 
grand canonical ensemble). Combining  Eqs.~(\ref{rhofin}), (\ref{r1fin}) and 
(\ref{idrho}), we obtain
\be
\frac{\rho^2}{s^2} h^{(2)}(\x,\alpha;\y,\gamma) = \frac{\rho}{s}
\frac{\dd [B(\x)\p(\alpha)]}{\delta [h(\y) \p (\gamma)]}\Bigg\vert_{h=0, B=0} 
- \frac{\rho}{s}\delta (\x - \y)\, \delta_{\alpha, \gamma}
\label{hrs}
\ee
Since the system is now homogeneous, $h^{(2)}$ depends only on the difference
$\x -\y$. Taking the Fourier transform of Eq.~(\ref{hrs}), we have that
\be
\frac{\rho^2}{s^2} \hat{h^{(2)}}(\vec{k};\alpha,\gamma) = - \frac{\rho}{s}
\delta_{\alpha, \gamma} +
\frac{\rho}{s}\int d \vec{k} \, \,e^{ i \vec{k}(\x-\y)} \,\frac{\dd [B(\x)
\p(\alpha)]}{\delta [h(\y) \p (\gamma)]}\Bigg\vert_{h=0, B=0} 
\label{hft}
\ee
Finally, we note that from the definition of the generalized functional 
derivative,
\be
\frac{\dd [B(\x)\p(\alpha)]}{\delta [h(\y) \p (\gamma)]} = \frac{\delta B(\x)}
{\delta h(\y)} \delta_{\alpha,\gamma}
\ee
where the expression $\delta B / \delta h$ on the r.h.s is now a usual 
functional derivative.The properties of the usual functional derivative imply 
that
\be
\int d \vec{k} \, e^{ i \vec{k}(\x-\y)} \frac{\delta B(\x)}
{\delta h(\y)}\Bigg\vert_{h=0, B=0} = \frac{\delta \hat{B}(\vec{k})}
{\delta \hat{h}(\vec{k})}\Bigg\vert_{h=0, B=0} = 
\frac{1}{1 - (\rho/s)\hat{f^{\dagger}}(\vec{k})}
\ee
where in the last equality we have used Eq.~(\ref{bfs}). Substituting this 
result in Eq.~(\ref{hft}) finally yields
\be
\hat{h^{(2)}}(\vec{k};\alpha,\gamma) = \frac{\hat{f^{\dagger}}(\vec{k})}
{1 - (\rho/s)\hat{f^{\dagger}}(\vec{k})} \delta_{\alpha,\gamma}
\label{hfin}
\ee
Hence, substituting this into Eq.~(\ref{gdagf}) finally yields the pair
connectedness within the mean field approximation,
\be
\hat{\gd}(\vec{k}) = \frac{ \hat{f^{\dagger}}(\vec{k})}
{1 - \rho\hat{f^{\dagger}}(\vec{k})}
\label{gdmf}
\ee
This expression is consistent with Eq.~(\ref{mfres}) for the mean cluster size
through the relation between $S$ and $\gd$ given by Eq.~(\ref{Stwo}).

Since $f^{\dagger}(\r) = f^{\dagger}(-\r)$, then for small $\vec{k}$, we have 
that
\be
\hat{f^{\dagger}}(\vec{k}) = \ve - \dve k^2 + O(k^4)
\ee
where
\bea
\ve &=& \int d \r f^{\dagger}(r) \nonumber \\
\dve &=& \int d \r \,\, \frac{r^2}{6} f^{\dagger}(r)
\eea
Therefore, for small wavevectors, the pair-connectedness is
\be
\hat{\gd}(\vec{k}) \approx \left(\frac{\hat{f^{\dagger}}(\vec{k})}
{1 - \rho \ve} \right) \frac{1}{1 + [\dve/(1 - \rho \ve)]k^2}
\ee
We expect by analogy with other critical phenomena that 
\be
\hat{\gd}(\vec{k}) \sim \frac{1}{1 + \xi^2k^2}
\ee
where $\xi$ is the {\it connectedness length}, which plays in percolation 
theory a role analogous the the one the correlation length plays in other
critical phenomena \cite{stauffer}. Hence, we have obtained an expression for 
the connectedness length within mean field theory,
\be
\xi^2 = \frac{\dve}{1 - \rho \ve} = \frac{\dve \rho_c}{\rho_c - \rho} \sim
(\rho_c - \rho)^{- \nu}
\ee
Thus, we find that the mean field value of the exponent $\nu$ is
\be
\nu = \frac{1}{2} \qquad \mbox{(Mean Field)} \label{nu}
\ee

\section{Discussion}
\setcounter{equation}{0}

The results of the mean field theory are summarized below.
\bea
P &=& 1 - \exp\left(- \rho P \ve \right) \nonumber \\
S &=& \frac{1}{1 - \rho \ve} \qquad\qquad (\rho < \rho_c) \nonumber \\
\xi^2 &=& \frac{\dve}{1 - \rho \ve}  \nonumber \\
\rho_c &=& \frac{1}{\ve} = \frac{1}{\int d\r f^{\dagger}(\r)} 
\nonumber \\
\beta_{MF} &=& 1 \nonumber \\
\gamma_{MF}&=&1  \nonumber \\
\nu_{MF} &=& \frac{1}{2}  
\eea

The critical exponents obtained are identical to those obtained in percolation 
on a Bethe lattice, usually considered as the mean field equivalent for the 
lattice percolation model. This therefore addresses one of the questions
presented in the introduction of I, namely the extent and nature of the
continuum percolation universality class. As far as mean field theory is 
concerned, we have obtained theoretical proof that this universality class
encompasses all interactions $v(i,j)$ and all binding criteria $p(i,j)$ and 
that it is identical to the universality class of lattice percolation. This 
last, incidentally, can be considered as a particular type of continuum 
percolation with interactions fixing the particles on the sites of a lattice. 
Therefore, the question of the independence of the critical exponents on the
interaction is connected essentially to the relation between continuum and
lattice percolation. We know, however, that mean field theory always grossly 
overestimates universality, as demonstrated by its being insensitive even to 
spatial dimensionality. Thus one cannot give too much weight to these 
conclusions. Nevertheless, the mean field result is an encouraging first step. 
Certainly had we found any difference between continuum and lattice percolation
in the mean field approximation, or any dependence on the interaction or the 
binding criterion, it would have been highly improbable that a more elaborate 
calculation would have restored universality. Thus at least, our results are 
encouraging, if far from definitive.

As to the critical density, we do not expect it to be quantitatively adequate,
as mean field results never are. However, as the dimensionality of the system 
increases, these results become better and better. Thus, Alon et al. 
\cite{alon} have shown that indeed,  for a system of permeable hyper-cubes,
$\rho_c V_{exc} \to 1$ for high dimensions. 

As another example, for a $D$-dimensional system of hard-cores and 
soft-shells \cite{bug, des}, we have
\be
v(i,j) = \left\{ \begin{array}{r @{\quad} l}
            \infty & \mbox{if \quad}\vert \r_i - \r_j \vert < a \\ 
              0 &  \mbox{otherwise}\end{array} \right. 
\ee

\be
p(i,j) = \left\{ \begin{array}{r @{\quad} l}
            \infty & \mbox{if \quad}\vert \r_i - \r_j \vert < d \\ 
             0 &  \mbox{otherwise}\end{array} \right.
\ee
where $a$ and $d$ are two length parameters. Then we obtain in the mean field
approximation
\be
\rho_c = \frac{1}{C_D \left( d^D - a^D \right)} =  \frac{1}{C_D d^D \left( 1 - 
\eta^D \right)}
\ee
where $\eta \equiv \a / d$  is the aspect ratio, and $C_D$ id the 
$D$-dimensional volume of a unit sphere. Hence, in mean field theory, $\rho_c$
is a monotonic function of the aspect ratio $\eta$. This 
contradicts the result of simulations which show $\rho_c$ to have a pronounced
minimum in 3 dimensions. However, some computer simulations \cite{phd} show
that as the dimensionality increases, the minimum becomes weaker, and at 
high dimensions, it appears that $\rho_c$ does increase ever more 
monotonically. Hence the mean field result corresponds again to the limit 
$D \to \infty$.

The importance of mean field theory is thus not so much in the numerical 
results is yields, but rather in the proof of the usefulness of the theoretical
approach presented here. The mean field equations have been derived wholly
within the Potts fluid picture, where heavy use is made of the available
Hamiltonian formulation. The $s \to 1$ limit then yields the percolation
quantities, which do indeed have the expected mean field properties. It is hard
to see how such results could have been obtained directly in the percolation 
picture. This indicates that other methods which can be applied to the Potts
fluid should yield interesting results for the continuum percolation problem.
This will be the subject of future inquiries.

\section*{Appendix}
\setcounter{equation}{0}
\renewcommand{\theequation}{A.\arabic{equation}}

The functional $F_t[B]$ is defined as in Eq.~(\ref{ft})
\bea
F_t(B) &\equiv& \frac{1}{N}\ln Z_0 + \frac{(N-1)}{2}\left\langle 
q(i,j)\exp\left[-\beta v(i,j)\right] - 1\right\rangle_0 \nonumber \\
&+& \frac{(N-1)}{2}\left\langle  p(i,j)
\exp\left[-\beta v(i,j)\right]\delta_{\la_i,\la_j}\right\rangle_0
\nonumber \\
&+& \left\langle \beta \left[h(j) -B(j)\right] \p (\la_j) 
\right\rangle_0 
\label{aft}
\eea

We now have that
\be
\ln Z_0 = N\ln(\G) - \ln(N!)
\ee
where
\bea
\G \equiv \sum_{\alpha} \int d\r \, \exp\left[\beta B(\r) \p(\alpha)\right]
\label{gam}
\eea
Similarly, we have immediately that
\be
\left\langle \beta \left[h(j) -B(j)\right] \p (\la_j) \right\rangle_0 =
\frac{1}{\G}\sum_{\la_j} \int dj \, e^{\beta B(j) \p(\la_j)}\left[h(j) -B(j)
\right] \p (\la_j) 
\ee
And
\bea
\lefteqn{\left\langle q(i,j)\exp\left[-\beta v(i,j)\right]\right\rangle_0} 
\nonumber \\
&=&\frac{1}{\G^2}\sum_{\la_i, \la_j} \int d i\, d j \,\,q(i,j)e^{-\beta v(i,j)}
e^{\beta [B(i) \p(\la_i)+ B(j) \p(\la_j)]} \\
\lefteqn{\left\langle  p(i,j)\exp\left[-\beta v(i,j)\right]\delta_{\la_i,\la_j}
\right\rangle_0} \nonumber \\
&=& \frac{1}{\G^2}\sum_{\la_i, \la_j}\int d i \,d j \,\,p(i,j)e^{-\beta v(i,j)}
e^{\beta [B(i) \p(\la_i)+ B(j) \p(\la_j)]}\delta_{\la_i,\la_j}
\eea
Taking now the functional derivatives of these terms, we find
\be
\frac{\dd\ln Z_0}{\delta[N\beta B(\x)\p (\la)]} = \frac{1}{\G}\exp\left[\beta 
B(\x) \p(\la)\right]
\ee
\bea
\lefteqn{\frac{\dd\left\langle \beta \left[h(j) -B(j)\right] \p (\la_j) 
\right\rangle_0}{\delta[\beta B(\x)\p (\la)]}} \nonumber \\
&=& \iex \frac{1}{\G}\left\{ -1 + \left[h(\x) -B(\x)\right] \p (\la)
\right\} \nonumber \\
&-&\iex\frac{1}{\G^2}\sum_{\la_j} \int dj \, e^{\beta B(j) \p(\la_j)}
\left[h(j) -B(j)\right] \p (\la_j) \\
\lefteqn{\frac{\dd\left\langle q(i,j)\exp\left[-\beta v(i,j)\right] - 1 
\right\rangle_0}{\delta[\beta B(\x)\p (\la)]}}
\nonumber \\
&=&\iex\frac{2}{\G^2}\sum_{\la_i} \int d \y \, q(\x -\y)
e^{-\beta v(\x -\y)}e^{\beta [B(\y) \p(\la_i)}  \nonumber \\
&-&\iex\frac{2}{\G^3}\sum_{\la_i, \la_j} \int di\, d j \,\, q(i,j)
e^{-\beta v(i,j)} \, e^{\beta [B(i) \p(\la_i)+ B(j) \p(\la_j)]} \\
\lefteqn{\frac{\dd\left\langle  p(i,j)\exp\left[-\beta v(i,j)\right]
\delta_{\la_i,\la_j}\right\rangle_0}{\delta[\beta B(\x)\p (\la)]} }
\nonumber \\
&=& \iex\frac{2}{\G^2}\int d \y \,p(\x -\y)e^{-\beta 
v(\x -\y)}e^{\beta [B(\y) \p(\la_i)}  \nonumber \\
&-&\iex\frac{2}{\G^3}\sum_{\la_i, \la_j} \int d i \, d j \,\, p(i,j)
e^{-\beta v(i,j)}e^{\beta [B(i) \p(\la_i)+ B(j) \p(\la_j)]}\delta_{\la_i,\la_j}
\nonumber \\
\eea

We now expand these expressions around $B \to 0$. We have that
\be
\G = \sum_{\alpha} \int d\r \, \exp \left[ \beta B(\r) \p(\alpha)\right] 
\approx \sum_{\alpha} \int d\r \left[ 1 + \beta B(\r) \p(\alpha)\right]
\ee
We note that
\be
\sum_{\alpha} \p(\alpha) = (s-1) + (s-1)(-1) = 0
\label{idipsi}
\ee
Therefore, to first order,
\be
\G = s V + O(B^2)
\ee
The identity Eq.~(\ref{idipsi}) simplifies greatly all the functional 
derivatives. To first order,
\bea
\frac{\dd\ln Z_0}{\delta[N\beta B(\x)\p (\la)]} &=& \frac{1}{sV}
\exp\left[\beta B(\x) \p(\la)\right]  + O(B^2) \\
\frac{\dd\left\langle q(i,j)\exp\left[-\beta v(i,j)\right] - 1 
\right\rangle_0}{\delta[\beta B(\x)\p (\la)} &=& O(B^2) 
\eea
And
\bea
\lefteqn{\frac{\dd\left\langle \beta \left[h(j) -B(j)\right] \p (\la_j) 
\right\rangle_0}{\delta[\beta B(\x)\p (\la)]}} \nonumber \\
&=& \iex \frac{1}{sV}\left\{ -1 + \left[h(\x) -B(\x)\right] \p (\la)
\right\} + O(B^2) \\
\lefteqn{\frac{\dd\left\langle  p(i,j)\exp\left[-\beta v(i,j)\right]
\delta_{\la_i,\la_j}\right\rangle_0}{\delta[\beta B(\x)\p (\la)]}} 
\nonumber \\
&=& \iex \frac{2}{s^2V^2}\int d \y \, p(\x -\y)e^{-\beta 
v(\x -\y)}B(\y) + O(B^2)
\eea
 
Summing up all these expressions and equating $\dd F_t /\delta [\beta B \p]=0$,
we obtain the condition
\be
B(\x) = h(\x) + \frac{\rho}{s}\int d \y \, p(\x- \y)
e^{-\beta v(\x -\y)} B(\y) + O(B^2)
\ee

\end{document}